\documentclass{article}
\usepackage{ijcai26}

\usepackage{times}
\usepackage{soul}
\usepackage{url}
\usepackage[hidelinks]{hyperref}
\usepackage[utf8]{inputenc}
\usepackage[small]{caption}
\usepackage{graphicx}
\usepackage{amsmath}
\usepackage{amsthm}
\usepackage{booktabs}
\usepackage{algorithm}
\usepackage{algorithmic}
\usepackage{multirow}
\usepackage{subcaption}
\usepackage{amssymb}
\usepackage{enumitem}
\usepackage[switch]{lineno}

\title{NiMark: A Non-intrusive Watermarking Framework against Screen-shooting Attacks}

\iftrue
\author{
Yufeng Wu$^1$
\and
Xin Liao$^1$\and
Baowei Wang$^2$\and
Han Fang$^3$\and
Xiaoshuai Wu$^1$\And
Guiling Wang$^4$
\affiliations
$^1$College of Cyber Science and Technology, Hunan University\\
$^2$School of Computer Science, Nanjing University of Information Science and Technology\\
$^3$School of Computing, National University of Singapore\\
$^4$Department of Computer Science, New Jersey Institute of Technology
\emails
wuyufeng0523@163.com, xinliao@hnu.edu.cn, wbw.first@163.com,
fanghan@nus.edu.sg, shinewu@hnu.edu.cn, gwang@njit.edu
}

\begin{document}

\maketitle

\begin{abstract}
Unauthorized screen-shooting poses a critical data leakage risk. Resisting screen-shooting attacks typically requires high-strength watermark embedding, inevitably degrading the cover image. To resolve the robustness-fidelity conflict, non-intrusive watermarking has emerged as a solution by constructing logical verification keys without altering the original content. However, existing non-intrusive schemes lack the capacity to withstand screen-shooting noise. While deep learning offers a potential remedy, we observe that directly applying it leads to a previously underexplored failure mode, the Structural Shortcut: networks tend to learn trivial identity mappings and neglect the image-watermark binding. Furthermore, even when logical binding is enforced, standard training strategies cannot fully bridge the noise gap, yielding suboptimal robustness against physical distortions. In this paper, we propose NiMark, an end-to-end framework addressing these challenges. First, to eliminate the structural shortcut, we introduce the Sigmoid-Gated XOR (SG-XOR) estimator to enable gradient propagation for the logical operation, effectively enforcing rigid image-watermark binding. Second, to overcome the robustness bottleneck, we devise a two-stage training strategy integrating a restorer to bridge the domain gap caused by screen-shooting noise. Experiments demonstrate that NiMark consistently outperforms representative state-of-the-art methods against both digital attacks and screen-shooting noise, while maintaining zero visual distortion.
\end{abstract}

\section{Introduction}

\begin{figure}[!t]
\centering
\includegraphics[width=0.48\textwidth]{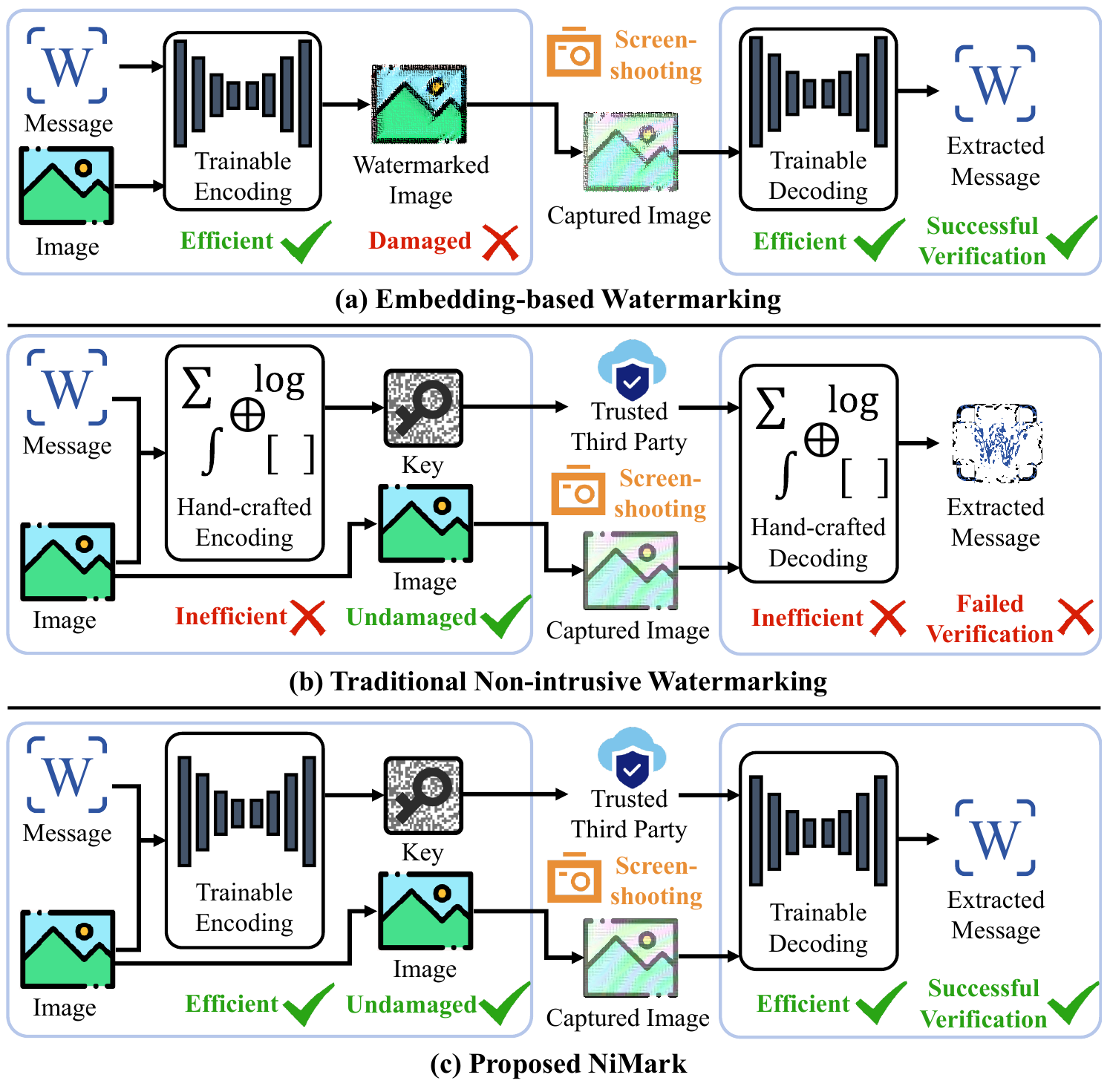}
\caption{Comparison of different watermarking frameworks against screen-shooting attacks.
(a) Embedding-based Watermarking: Although robust against screen-shooting, it inevitably degrades the cover image quality by explicitly modifying pixel values.
(b) Traditional Non-intrusive Watermarking: While preserving the original image, it fails to resist the severe distortions caused by screen-shooting due to the limited robustness of hand-crafted features.
(c) Proposed NiMark: By harnessing the powerful capability of deep learning, our framework achieves strong robustness against screen-shooting attacks within a non-intrusive architecture, guaranteeing zero visual distortion.}
\label{fig:first}
\end{figure}
The illicit capture of displayed content via external cameras constitutes a severe threat to information security. Digital watermarking has long served as a fundamental defense mechanism for identifying ownership and tracing leakage sources. However, unlike lossless digital transmission, this process introduces complex distortions like Moir\'e patterns and lightness distortion \cite{fang2018screen,wengrowski2019light}, rendering most conventional watermarking schemes ineffective.
Current solutions predominantly rely on embedding-based watermarking \cite{tancik2020stegastamp,fang2022pimog,li2024screen}. While robust, these methods inevitably degrade visual quality by modifying pixel values (Fig.~\ref{fig:first}(a)). In high-security applications requiring strict data integrity, such modifications are prohibitive, necessitating non-intrusive alternatives.

To circumvent the fidelity-robustness trade-off, non-intrusive watermarking, also known as zero-watermarking \cite{wen2003concept}, has emerged as a vital research direction. Fig.~\ref{fig:first}(b) depicts this approach, where instead of modifying the image, the algorithm extracts essential features and logically combines them with the copyright message to construct a verification key. However, the current state of non-intrusive watermarking research faces a significant bottleneck. The vast majority of existing schemes still rely on traditional, hand-crafted algorithms. These manual feature extractors are designed based on prior knowledge of signal processing and lack the adaptive capacity to capture deep, semantic features resilient to the severe non-linear distortions of screen-shooting. Consequently, as shown in the middle row of Fig.~\ref{fig:first}, traditional methods often fail to verify copyright under real-world screen-shooting attacks.

Conversely, while deep neural networks have revolutionized image processing, an end-to-end deep learning framework for robust non-intrusive watermarking remains a virtually untouched domain. In exploring this uncharted territory, we identify a critical obstacle, the Structural Shortcut. Non-intrusive watermarking necessitates a strong binding between the cover image and the copyright message. However, enforcing this dependency in a deep learning framework is challenging. Lacking a mechanism to enforce logical binding, the network may converge to a degenerate solution that largely neglects the image features and instead approximates a trivial identity mapping. Crucially, if the key becomes a self-contained carrier independent of the image, the copyright binding is severed, undermining the protection mechanism.

Furthermore, even if the structural shortcut is averted, a secondary challenge persists: the Noise Domain Gap. Simultaneously optimizing for logical binding and resistance to screen-shooting complicates the learning process. Standard training strategies cannot fully bridge the domain gap, resulting in suboptimal robustness against real-world noise.

To address these challenges, we propose NiMark, the first end-to-end deep learning framework tailored for screen-shooting resistant non-intrusive watermarking, as visualized in Fig.~\ref{fig:first}(c). Unlike previous approaches, NiMark utilizes deep neural networks to ensure robustness while guaranteeing zero visual distortion. Our methodology is designed to mitigate the structural shortcut and achieve robustness through model construction and a tailored training strategy.

First, drawing inspiration from traditional non-intrusive schemes, we identify the logical XOR operation as the essential constraint for binding the image features and the watermark. To integrate this discrete logic into a deep learning framework, we propose the Sigmoid-Gated XOR (SG-XOR) Estimator. By formulating a differentiable proxy for the discrete XOR logic, this module imposes a constraint during backpropagation. This design significantly increases the dependency of the generated key on the cover image features, making trivial identity mappings empirically unfavorable during optimization.

Second, to achieve robustness against screen-shooting, we devise a two-stage training strategy. Recognizing that screen-shooting noise creates a severe domain gap, we integrate a dedicated Restorer into the pipeline. By decoupling the logical binding learning from the physical restoration, this strategy ensures that the model first learns the correct dependency and then adapts to the screen-shooting noise by recovering the image quality before decoding.

The main contributions of this paper are summarized as follows:
\begin{itemize}
    \item We propose NiMark, the first end-to-end deep learning framework for non-intrusive watermarking robust against screen-shooting, effectively overcoming the bottlenecks of traditional hand-crafted features.
    \item We design the SG-XOR Estimator to enable gradient propagation through the discrete logical generation. This module enforces image-watermark binding by discouraging the network from converging to trivial structural shortcuts.
    \item We introduce a two-stage training strategy that integrates an independent restorer. This approach bridges the screen-shooting domain gap by rectifying distortions prior to decoding.
    \item Experiments demonstrate that NiMark achieves consistently stronger robustness over state-of-the-art methods, while maintaining zero visual distortion under the non-intrusive setting.
\end{itemize}

\section{Related Work}
\label{sec:related}

\subsection{Embedding-based Robust Watermarking}
Traditional watermarking techniques operate in spatial or transform domains \cite{wolfgang2002perceptual,chu2003dct,zhang2020image,liu2002svd} but struggle with non-linear distortions. Deep learning has transformed this field \cite{wu2025robust}, with end-to-end frameworks like HiDDeN \cite{zhu2018hidden} and subsequent refinements \cite{liu2019novel,jia2021mbrs,sander2025watermark}.
Specifically for the screen-camera channel, early template matching \cite{fang2018screen,fang2021tera} has been superseded by deep methods employing differentiable noise layers. StegaStamp \cite{tancik2020stegastamp} and PIMoG \cite{fang2022pimog} simulate perspective and Moir\'e distortions, while others explore inverse encoding \cite{wang2025moire}. Recent works address geometric synchronization \cite{he2024camera}, partial occlusion \cite{ma2025ropass,chen2025flexible}, and data-driven simulation \cite{wengrowski2019light,gao2025screen,wu2026sim}. However, these methods inherently modify the cover image pixels to embed signals, which compromises visual fidelity and creates a bottleneck for applications requiring strict data integrity.

\subsection{Non-Intrusive Watermarking and Shortcut Learning}
Non-intrusive schemes \cite{wen2003concept} construct logical verification keys to preserve image quality. Traditional algorithms rely on orthogonal moments \cite{gao2015bessel,wang2017robust} or their geometric invariants \cite{wang2016geometrically,wang2019ternary,yang2020color,yang2021robust}. Recent studies have extended these to light fields \cite{wen2025lfizw}, Radon space \cite{qi2024representing}, and blockchain-based protocols \cite{wang2022image,hou2025zero}. Despite ensuring zero distortion, these hand-crafted features lack robustness against severe screen-shooting attacks.

While deep learning approaches have emerged \cite{li2025dual,liu2025attack}, applying them to this task risks triggering shortcut learning \cite{geirhos2020shortcut}. In general vision tasks, models tend to exploit superficial correlations such as background \cite{beery2018recognition}, texture \cite{geirhos2018imagenet}, or high-frequency artifacts \cite{jo2017measuring,smeu2025circumventing}. However, the manifestation of this phenomenon in non-intrusive watermarking remains unexplored. In this work, we identify a domain-specific failure mode termed the Structural Shortcut, where the network completely bypasses the image-watermark binding to minimize loss. This critical vulnerability, which NiMark aims to resolve, represents a unique challenge distinct from previously studied shortcuts.

\begin{figure*}[t]
\centering
\includegraphics[width=\textwidth]{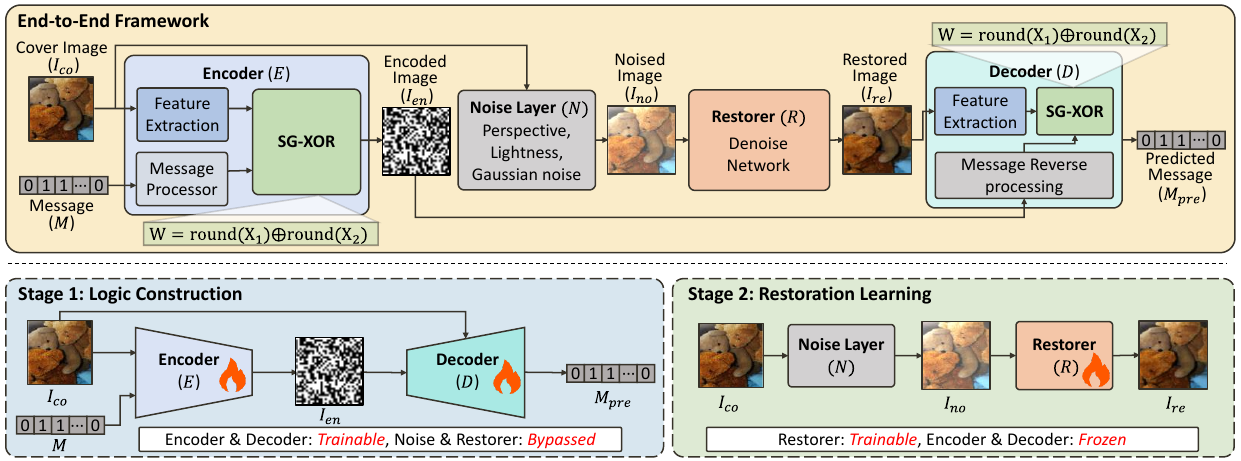} 
\caption{Overview of the proposed NiMark framework. The top panel illustrates the detailed network architecture which includes the Encoder integrated with SG-XOR, the Noise Layer, the Restorer, and the Decoder. The bottom panels demonstrate the proposed two-stage training strategy. Stage 1 constructs the logical dependency in a noise-free environment. Stage 2 independently optimizes the restoration network to handle screen-shooting distortions.}
\label{fig:framework}
\end{figure*}

\section{Methodology}
\label{sec:methodology}

Fig.~\ref{fig:framework} presents the NiMark framework. This section details four core components of the proposed watermarking system: the network architecture, the SG-XOR Estimator for enforcing logical binding, and a two-stage training strategy for handling screen-shooting distortions.

\subsection{Network Architecture}
The overall framework, shown in the top panel of Fig.~\ref{fig:framework}, consists of four sequential modules: Encoder, Noise Layer, Restorer, and Decoder.

\subsubsection{Encoder}
The goal of the encoder is to generate a key from the cover image $I_{co} \in \mathbb{R}^{C \times H \times W}$ and a binary message $M \in \{0,1\}^L$.
To facilitate the interaction between the 1D message vector and the 2D image features, we design a dual-branch architecture.
For the image feature extraction, we employ a differentiable moment feature extractor. We project the grayscale version of $I_{co}$ onto a set of radial harmonic kernels defined in the polar coordinate system to directly extract a compact geometric feature map.
Simultaneously, for the message branch, we employ a Message Processor. In this module, the message vector is first reshaped and then processed using convolution layers to match the spatial resolution of the extracted image features.
Let $E(\cdot)$ denote the encoder. It fuses the normalized moment features with the processed message features to generate the encoded key $I_{en} \in \{0, 1\}^{C' \times H' \times W'}$, where $C', H', W'$ denote the channel, height, and width of the feature maps, respectively.
\begin{equation}
I_{en} = E(I_{co}, M)
\end{equation}

\subsubsection{Noise Layer}
In deep learning-based watermarking, the noise layer acts as a critical adversary during training. By accurately simulating real-world distortions, it drives the network to adaptively learn robust representations that can survive such attacks. Although our framework is non-intrusive, we follow this established strategy to acquire robustness. We integrate a differentiable noise layer $N(\cdot)$ to mimic the analog-to-digital screen-shooting process, comprising four key physical distortions:
\noindent \textbf{a) Perspective Distortion:} We apply a random 4-point perspective transformation matrix to simulate geometric deformation. \textbf{b) Lightness Distortion:} A linear brightness shift is applied defined as $I' = I + b$, where $b \sim U(0.5, 0.7)$. \textbf{c) Moir\'e Pattern:} We inject Moir\'e distortions generated using superimposed cosine functions with random orientations ($p=0.2$). \textbf{d) Gaussian Noise:} Additive noise is introduced to simulate sensor grain.

In the non-intrusive setting, these distortions are applied directly to the cover image, denoted as $I_{no} = N(I_{co})$. It is worth noting that this module serves as a flexible component, which can be interchanged with distinct noise layers to target different attack scenarios.

\subsubsection{Restorer}
The Restorer, denoted as $R(\cdot)$, acts as a crucial preprocessing module to mitigate the domain gap. It aims to rectify the noise in the attacked image $I_{no}$ to recover the visual content of the original cover. This restoration process is mathematically formulated as:
\begin{equation}
I_{re} = R(I_{no})
\end{equation}
where $I_{re}$ represents the restored image. Ideally, $R(\cdot)$ learns the inverse transformation of the noise layer such that $I_{re} \approx I_{co}$, providing a cleaner reference for the subsequent decoding step.

\subsubsection{Decoder}
The Decoder $D(\cdot)$ extracts the watermark. In our non-intrusive setting, decoding is key-dependent. The decoder takes the generated encoded image $I_{en}$ and the restored reference image $I_{re}$ to retrieve the message:
\begin{equation}
M_{pre} = D(I_{en}, I_{re})
\end{equation}
The decoding process utilizes a Message Reverse Processing, which employs strided convolutions to downsample the feature maps back to the message dimension $L$.

\subsection{Sigmoid-Gated XOR Estimator}
\label{sec:ste_xor}

\begin{figure*}[t]
\centering
\begin{subfigure}[b]{0.24\textwidth}
\centering
\includegraphics[width=\linewidth]{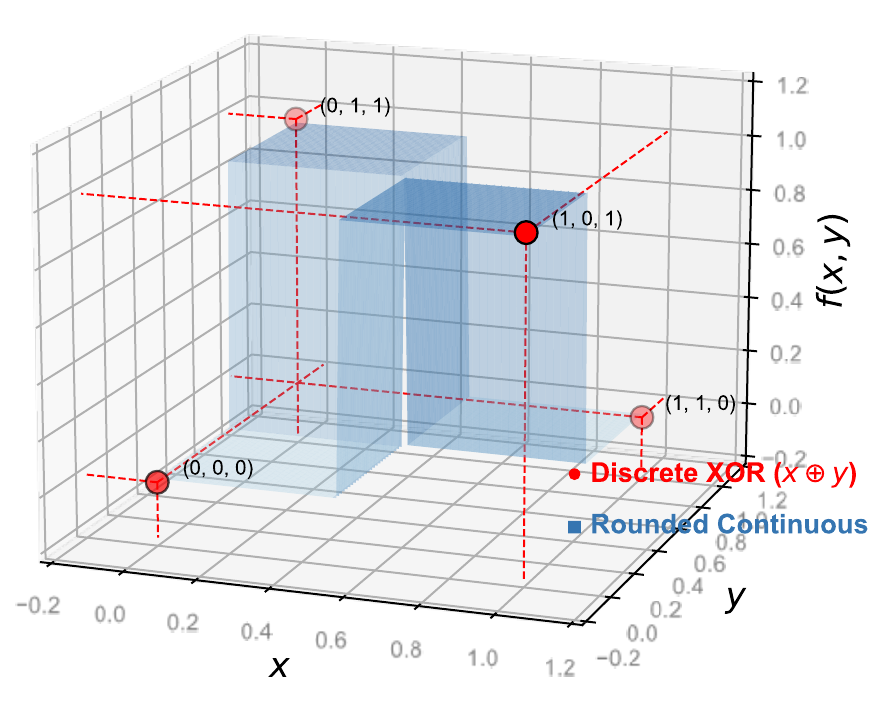}
\caption{Non-Differentiable Function}
\label{fig:piecewise_target}
\end{subfigure}
\hfill
\begin{subfigure}[b]{0.24\textwidth}
\centering
\includegraphics[width=\linewidth]{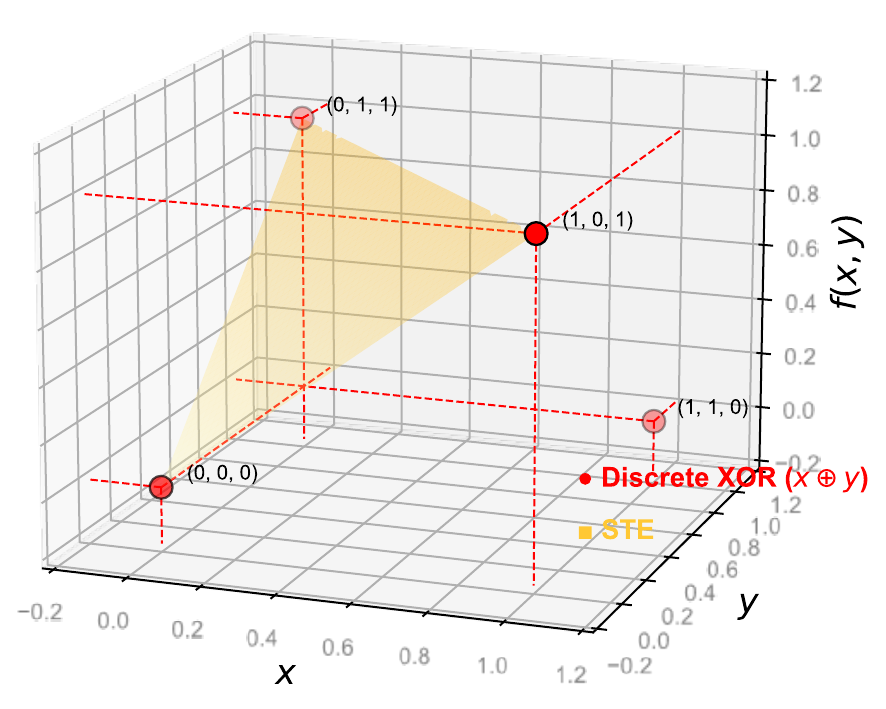}
\caption{STE Proxy}
\label{fig:ste_proxy}
\end{subfigure}
\hfill
\begin{subfigure}[b]{0.24\textwidth}
\centering
\includegraphics[width=\linewidth]{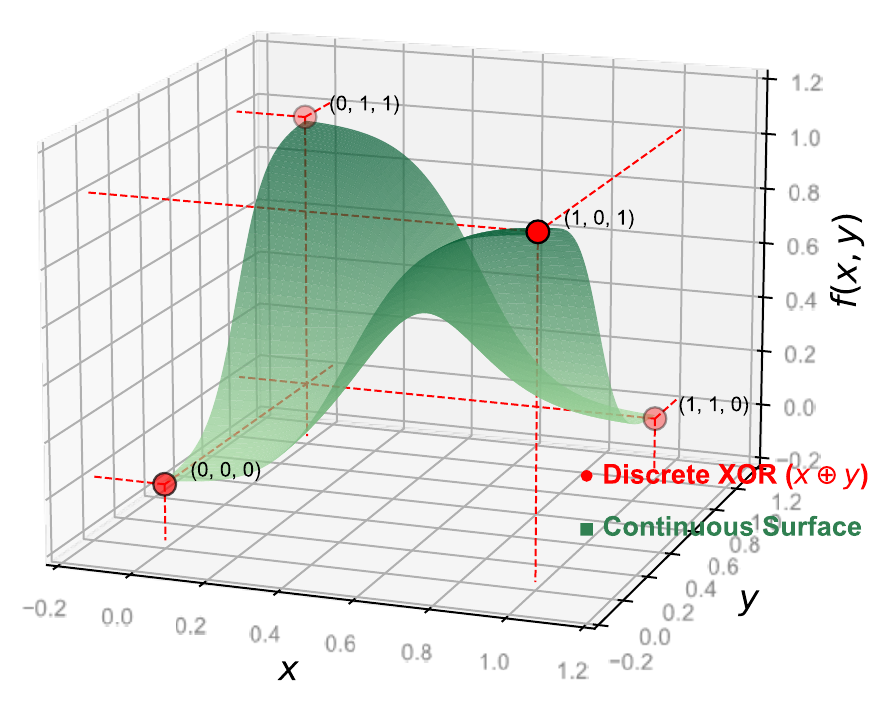}
\caption{Proposed SG-XOR Proxy}
\label{fig:sg_xor_proxy}
\end{subfigure}
\hfill
\begin{subfigure}[b]{0.24\textwidth}
\centering
\includegraphics[width=\linewidth]{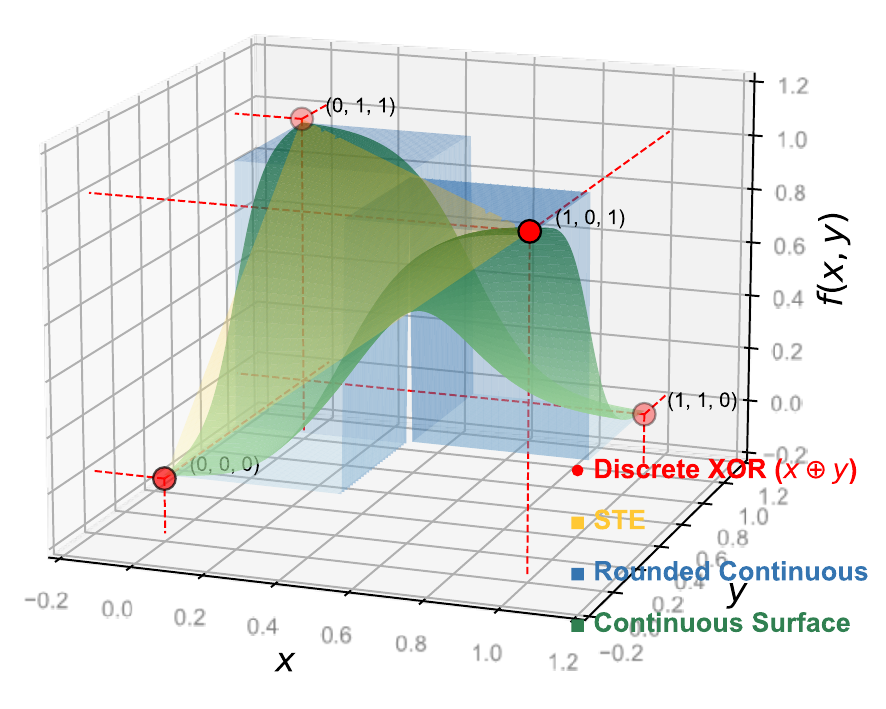}
\caption{Comparison of All Functions}
\label{fig:comparison_all}
\end{subfigure}

\caption{Visualization of the discrete XOR operation and its various differentiable proxies for gradient estimation. (a) The true, non-differentiable surface of the compound operation $W = \text{round}(X_1) \oplus \text{round}(X_2)$, which defines the discrete target behavior and causes gradient cliffs that prevent standard backpropagation. (b) The standard linear STE proxy, demonstrating a crude and misaligned gradient approximation. (c) The proposed SG-XOR proxy, providing a smooth, differentiable manifold that aligns well with the discrete XOR logic while maintaining differentiability. (d) Direct comparison among the SG-XOR proxy (green), STE proxy (yellow), and target discrete function (blue), showing the superior consistency of our approach.}
\label{fig:xor_visualization}
\end{figure*}

A core component of our NiMark framework involves a hard binarization step using round($\cdot$) followed by a bitwise XOR operation. We explicitly design this module to force the model to learn a content-dependent watermark embedding. By inextricably linking the binarized image features with the binarized message features via XOR, the model is strongly encouraged to incorporate cover image features, as ignoring them leads to a clear performance degradation. This logical dependency explicitly reduces the tendency to converge to trivial structural shortcuts, ensuring a strong binding where the verification key is derived from the image features rather than being independent of the specific image content. The visual comparison of this discrete operation and our differentiable proxy is illustrated in Fig.~\ref{fig:xor_visualization}.

In the forward pass, this operation is formally defined by the compound discrete function:
\begin{equation}
W = \text{round}(X_1) \oplus \text{round}(X_2)
\end{equation}
where $X_1, X_2 \in [0, 1]^{C' \times H' \times W'}$ represent the pre-processed continuous feature tensors (feature maps). $\text{round}(\cdot)$ denotes the element-wise hard binarization function, and $\oplus$ is the element-wise logical XOR. The resulting discrete surface of $W$, as illustrated in Fig.~\ref{fig:piecewise_target}, exhibits sharp discontinuities and flat plateaus that block gradient flow, which poses a fundamental obstacle to gradient-based learning.

This operation is fundamentally non-differentiable. A common workaround, shown in Fig.~\ref{fig:ste_proxy}, is to use the Straight-Through Estimator (STE) \cite{bengio2013estimating}, which simply copies gradients from the output to the input. Effectively, it treats the binarization as an identity function. However, the standard STE, which typically approximates the gradient of a binarization function as an identity map, i.e., a gradient of 1, is ill-suited for this complex, compound operation. Such a crude, linear approximation bears little resemblance to the highly non-linear gradient of the true $\text{round}(\cdot) \oplus \text{round}(\cdot)$ logic. This severe mismatch between the discrete forward pass and the oversimplified backward pass leads to unstable training and significant performance degradation.

\begin{figure}[t]
\centering
\includegraphics[width=0.48\textwidth]{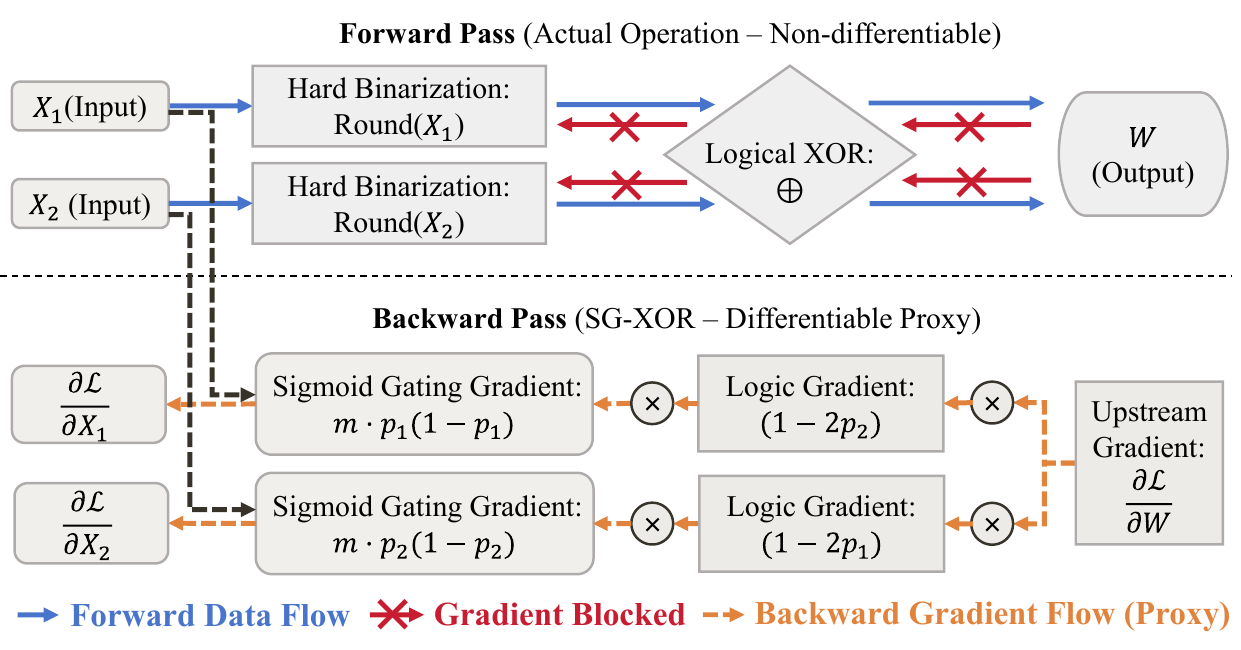} 
\caption{Mechanism of the proposed SG-XOR Estimator. The Forward Pass (top) performs the strict discrete operation $W = \text{round}(X_1) \oplus \text{round}(X_2)$, where gradients are naturally blocked by the hard binarization. The Backward Pass (bottom) utilizes our proposed differentiable proxy to estimate gradients. The upstream gradient is modulated by two distinct components: the Logic Gradient which enforces the XOR truth table and the Sigmoid Gating Gradient which stabilizes training near the decision boundaries.}
\label{fig:sg_xor}
\end{figure}

To resolve this, we design a specialized, differentiable proxy gradient that accurately models the target discrete behavior. We term our approach the SG-XOR Estimator. The backward pass is derived from a composite proxy function that models the binarization, or gating, and the XOR logic operations. As shown in the workflow diagram (Fig.~\ref{fig:sg_xor}), we seek to find the gradients $\partial \mathcal{L} / \partial X_1$ and $\partial \mathcal{L} / \partial X_2$ during backpropagation.

\subsubsection{Binarization Approximation}
First, we approximate the non-differentiable $\text{round}(\cdot)$ function with a parameterized sigmoid function $\sigma_{m,n}(\cdot)$:
\begin{equation}
p_1 = \sigma_{m,n}(X_1) = \frac{1}{1 + \exp(-m(X_1 + n))}
\label{eq:ste_sig_1}
\end{equation}
\begin{equation}
p_2 = \sigma_{m,n}(X_2) = \frac{1}{1 + \exp(-m(X_2 + n))}
\label{eq:ste_sig_2}
\end{equation}
where $p_1$ and $p_2$ represent the soft binarization of $X_1$ and $X_2$. This function is governed by two hyperparameters: the steepness parameter $m$ and the binarization threshold $n$. The parameter $m$ controls the sharpness of the sigmoid curve; we set $m=10$ to enforce a close approximation to a step function. The parameter $n$ defines the center of the transition. To accurately model the behavior of the round($\cdot$) function, which has its decision boundary at 0.5 for inputs in $[0, 1]$, we set the threshold to $n=-0.5$.

\subsubsection{Continuous XOR Approximation}
Next, we must define a continuous proxy for the discrete logic $p_1 \oplus p_2$. The target operation is defined by the XOR truth table. We select the bilinear polynomial that satisfies this condition, which is also equivalent to the probability of an XOR operation between two independent Bernoulli variables:
\begin{equation}
W_{proxy} = f(p_1, p_2) = p_1 + p_2 - 2p_1p_2
\label{eq:ste_proxy_f}
\end{equation}
This function aligns with the discrete XOR logic. For instance, at the input vertex $(1, 1)$, it yields $f(1, 1) = 1 + 1 - 2 = 0$, adhering to the truth table.

\subsubsection{Gradient Computation}
During the backward pass, we compute the gradient of the loss $\mathcal{L}$ with respect to the original inputs $X_1$ and $X_2$ by applying the chain rule through our composite proxy function $W_{proxy}$.
The detailed mathematical derivation and theoretical justification for this gradient flow are provided in the Supplementary Material.
Here, we directly present the final gradients utilized for backpropagation, which are composed of a Logic Gradient and a Sigmoid Gate:
\begin{equation}
\boxed{
\frac{\partial \mathcal{L}}{\partial X_1} = \frac{\partial \mathcal{L}}{\partial W_{proxy}} \cdot \underbrace{(1 - 2p_2)}_{\text{Logic Gradient}} \cdot \underbrace{(m \cdot p_1 (1 - p_1))}_{\text{Sigmoid Gate}}
}
\label{eq:ste_final_grad_1}
\end{equation}
\begin{equation}
\boxed{
\frac{\partial \mathcal{L}}{\partial X_2} = \frac{\partial \mathcal{L}}{\partial W_{proxy}} \cdot \underbrace{(1 - 2p_1)}_{\text{Logic Gradient}} \cdot \underbrace{(m \cdot p_2 (1 - p_2))}_{\text{Sigmoid Gate}}
}
\label{eq:ste_final_grad_2}
\end{equation}
where $p_1 = \sigma_{m,n}(X_1)$ and $p_2 = \sigma_{m,n}(X_2)$ are re-computed during the backward pass using the saved inputs $X_1$ and $X_2$. This formulation ensures both faithful approximation of the discrete logic and gradient-friendly learning dynamics.

\subsection{Two-Stage Training Strategy}
\label{sec:two_stage_training}

The specific data flows and component states for each training stage are visualized in the bottom panels of Fig.~\ref{fig:framework}.
Training deep networks directly under severe screen-shooting distortions is challenging due to the large domain gap.
We propose a two-stage training strategy that decouples logical binding learning from physical restoration, ensuring the model captures both the correct dependency and robustness against noise.

\subsubsection{Stage 1: Logic Construction}
The primary goal of this stage is to construct the correct feature-watermark dependency via SG-XOR in a noise-free environment.

In this first stage, we bypass the noise layer and the restorer. The goal is to train the Encoder ($E$) and Decoder ($D$) to perform the fundamental non-intrusive watermarking task. The training is guided solely by the message recovery loss $\mathcal{L}_{de}$, as the visual quality of the cover image is naturally preserved in our non-intrusive setting.
The decoder loss $\mathcal{L}_{de}$ measures the error between the predicted message $M_{pre}$ and the ground truth $M$:
\begin{equation}
\mathcal{L}_{stage1} = \mathcal{L}_{de} = \|M_{pre} - M\|_2^2
\end{equation}
where $\|\cdot\|_2^2$ denotes the Mean Squared Error (MSE). In this stage, the SG-XOR estimator enforces the logical binding. By learning to decode the message through the XOR operation, the network is forced to utilize the image features, thereby preventing the Structural Shortcut (identity mapping).

\subsubsection{Stage 2: Restoration Learning}
This stage aims to train the Restorer to rectify physical screen-shooting distortions independently.

In this second stage, we freeze the Encoder and Decoder. We focus exclusively on optimizing the Restorer ($R$). The network takes a distorted image and learns to recover the original cover:
\begin{equation}
I_{no} = N(I_{co}), \quad I_{re} = R(I_{no})
\end{equation}
The loss function $\mathcal{L}_{stage2}$ enforces restoration quality using MSE and MS-SSIM:
\begin{equation}
\mathcal{L}_{stage2} = \lambda_{pix} \cdot \|I_{re} - I_{co}\|_2^2 + \lambda_{str} \cdot (1 - \text{SSIM}(I_{re}, I_{co}))
\end{equation}
where we set $\lambda_{pix}=10$ and $\lambda_{str}=1$ to maintain consistent optimization objectives for pixel-level and structural fidelity. By recovering the image quality, the restorer effectively bridges the domain gap, allowing the pre-trained encoder and decoder from Stage 1 to function correctly even under attack conditions.

\begin{table*}[t]
\centering
\small
\setlength{\tabcolsep}{3pt}
\begin{tabular}{l|cc|cccc|cccc|cccccc}
\toprule
\multirow{3}{*}{Method} & \multicolumn{2}{c|}{Standard} & \multicolumn{4}{c|}{Horizontal Perspective: Left} & \multicolumn{4}{c|}{Horizontal Perspective: Right} & \multicolumn{6}{c}{Shooting Distance (Varying)} \\
\cmidrule{2-17}
 & \multicolumn{2}{c|}{($0^\circ$, 30cm)} & \multicolumn{2}{c}{$60^\circ$} & \multicolumn{2}{c|}{$30^\circ$} & \multicolumn{2}{c}{$30^\circ$} & \multicolumn{2}{c|}{$60^\circ$} & \multicolumn{2}{c}{40cm} & \multicolumn{2}{c}{50cm} & \multicolumn{2}{c}{60cm} \\
\cmidrule{2-17}
 & BER & U-BER & BER & U-BER & BER & U-BER & BER & U-BER & BER & U-BER & BER & U-BER & BER & U-BER & BER & U-BER \\
\midrule
RHFM   & 21.10 & 24.24 & 22.62 & 22.75 & 20.38 & 22.98 & 20.56 & 23.29 & 19.74 & 23.16 & 21.23 & 22.86 & 21.04 & 24.08 & 21.13 & 23.85 \\
EFM    & 19.24 & 20.40 & 19.91 & 20.67 & 18.23 & 20.95 & 18.64 & 20.76 & 18.55 & 20.36 & 19.01 & 19.97 & 19.44 & 20.60 & 19.63 & 21.22 \\
PCET   & 20.04 & 20.91 & 20.44 & 22.03 & 19.16 & 21.33 & 19.82 & 21.38 & 18.91 & 20.32 & 20.25 & 21.07 & 20.55 & 21.76 & 20.52 & 21.86 \\
BFM    & 20.51 & 22.66 & 21.90 & 22.85 & 19.82 & 21.85 & 20.18 & 22.40 & 19.85 & 23.45 & 20.70 & 23.53 & 20.42 & 23.54 & 21.16 & 22.60 \\
FJFM   & 21.83 & 23.62 & 22.55 & 23.45 & 21.34 & 23.75 & 21.81 & 24.32 & 21.37 & 24.02 & 22.51 & 23.76 & 21.92 & 23.63 & 22.40 & 22.92 \\
GPCET  & 19.26 & 20.69 & 19.39 & 20.93 & 18.25 & 20.70 & 19.05 & 20.28 & 17.96 & 20.50 & 19.57 & 21.09 & 19.77 & 20.42 & 19.66 & 21.23 \\
FJFMR  & 8.58  & 9.68  & 8.70  & 9.36  & 8.36  & 9.77  & 8.40  & 9.58  & 8.38  & 8.73  & 8.42  & 9.27  & 8.47  & 9.22  & 8.65  & 9.42  \\
GPCETR & 14.94 & 16.31 & 16.08 & 17.15 & 14.91 & 16.61 & 14.98 & 17.70 & 14.69 & 16.31 & 15.37 & 16.08 & 15.15 & 16.81 & 15.32 & 15.62 \\
GRHFMR & 8.72  & 14.77 & 13.30 & 14.51 & 12.58 & 14.25 & 12.59 & 13.93 & 12.26 & 13.96 & 12.85 & 13.64 & 12.52 & 14.20 & 12.87 & 13.75 \\
\midrule
\textbf{NiMark} & \textbf{0.46} & \textbf{49.41} & \textbf{1.13} & \textbf{53.82} & \textbf{0.97} & \textbf{40.25} & \textbf{0.68} & \textbf{48.16} & \textbf{1.28} & \textbf{47.32} & \textbf{0.57} & \textbf{51.98} & \textbf{0.58} & \textbf{51.64} & \textbf{0.59} & \textbf{51.03} \\
\bottomrule
\end{tabular}
\caption{Robustness (BER \%) and Binding Integrity (U-BER \%) under varying horizontal angles and shooting distances. The ``Standard'' column denotes the baseline setting ($0^\circ$, 30cm), while others vary a single parameter relative to this setting.}
\label{tab:ni_robustness}
\end{table*}

\section{Experimentation}
\label{sec:experiments}

In this section, we evaluate the performance of NiMark. We first detail the experimental setup, then present comparisons against state-of-the-art methods, and finally validate our design choices through ablation studies.

\subsection{Experimental Setup}

\subsubsection{Datasets and Implementation}
Adhering to the protocol in \cite{WU2024129282}, we used 1,000 COCO images for training (with two disjoint test sets of 500 and 100 images) and employed the AdamW optimizer with a weight decay of $0.02$ and a batch size of 16. Regarding the input configuration, consistent with \cite{wen2025lfizw}, we resized all images to $512 \times 512$ and set the message length to $L=1024$ bits, yielding an encoded key $I_{en}$ of size $1 \times 32 \times 32$. The framework was implemented in PyTorch on an NVIDIA RTX 3060 GPU. The training followed our proposed two-stage strategy for a total of 300 epochs, regulated by a cosine annealing scheduler with a warmup period.

\subsubsection{Baselines}
To evaluate performance, we selected a comprehensive set of baselines.

\noindent 1) Non-Intrusive Methods: We implemented 9 representative schemes based on orthogonal moments, including RHFM \cite{wang2019ternary}, EFM \cite{wang2016geometrically}, PCET \cite{wang2017robust}, BFM \cite{gao2015bessel}, FJFM \cite{yang2021robust}, GPCET \cite{yang2020color}, FJFMR \cite{qi2024representing}, GPCETR \cite{qi2024representing}, and GRHFMR \cite{wen2025lfizw}.

\noindent 2) Screen-Shooting Resilient Methods: We compared NiMark against leading embedding-based frameworks designed for the screen-camera channel, including StegaStamp \cite{tancik2020stegastamp}, PIMoG \cite{fang2022pimog}, SRWSS \cite{gao2025screen}, SSDS \cite{li2024screen} and S2R \cite{wu2026sim}.

\subsubsection{Evaluation Metrics}

\noindent 1) Robustness: Measured by Bit Error Rate (BER, \%). Lower is better.

\noindent 2) Visual Quality: Measured by PSNR (dB) and SSIM. For non-intrusive methods, the values are theoretically optimal (PSNR $\to \infty$, SSIM $= 1.0$).

\noindent 3) Unpaired BER (U-BER): To detect the Structural Shortcut, we verify whether the verification key is logically bound to the image content or has degenerated into a self-contained carrier. The U-BER evaluates the decoding performance when a specific valid key $I_{en}$ is paired with incorrect, uncorrelated reference images. Crucially, to ensure the metric reflects the model's behavior under attack conditions, the unpaired references undergo the exact same noise \textbf{and restoration} process as the correct cover. The U-BER for a single sample is calculated as the average error over $K$ random unpaired references:
\begin{equation}
\text{U-BER} = \frac{1}{K} \sum_{k=1}^{K} \left( \frac{1}{L} \sum_{i=1}^{L} \left| m_i - D(I_{en}, I'^{(k)}_{rand})_i \right| \right) \times 100\%
\end{equation}
\noindent where $K=10$ denotes the number of random reference images used for averaging to ensure stability, $L$ is the message length, $m_i$ is the $i$-th bit of the ground truth message, and $I'^{(k)}_{rand}$ represents the $k$-th restored random reference image. Ideally, $\text{U-BER} \approx 50\%$, indicating that the key alone is insufficient for decoding without the correct image features.

\subsection{Comparison with Non-Intrusive Schemes: Screen-Shooting Robustness}

Table \ref{tab:ni_robustness} presents the comprehensive evaluation results regarding horizontal perspective angles and varying shooting distances. Unless otherwise specified, experiments were conducted using a Lenovo Legion Y9000P display and a Samsung S20FE capture device, with a default setting of 30cm distance and a $0^\circ$ angle (perpendicular to the screen). Results for vertical angles (Up and Down) are provided in the Supplementary Material.

As shown in Table \ref{tab:ni_robustness}, traditional moment-based methods exhibit instability under varying distances, with BERs fluctuating between 19\% and 21\%. In contrast, NiMark demonstrates superior robustness, maintaining BERs consistently below 3.5\%. Crucially, NiMark achieves this while keeping U-BERs near 50\% across all tested conditions, confirming that the logical binding is robustly preserved even under severe geometric distortions.

\subsection{Comparison with Non-Intrusive Schemes: Digital Attacks}
To verify versatility, we also tested NiMark under standard digital attacks (JPEG, noise, filtering). Results provided in the Supplementary Material demonstrate that NiMark achieves competitive resilience, with U-BER consistently around 50\%, confirming rigid logical binding.

\subsection{Comparison with Screen-Shooting Resilient Schemes}
We further compare NiMark with representative screen-shooting resilient methods: StegaStamp, PIMoG, SRWSS, and S2R. Table \ref{tab:comparison_embedding} presents the results regarding robustness and visual quality. Most baselines typically operate on $128 \times 128$ images with a 30-bit payload ($\approx$ 0.0018 bpp). While S2R increases this to 64 bits ($\approx$ 0.0039 bpp), NiMark processes $512 \times 512$ inputs with a 1024-bit message, maintaining a comparable bit density. Experimental results demonstrate that NiMark achieves strong robustness against screen-shooting. Furthermore, as a non-intrusive framework, it guarantees zero visual distortion, offering a viable alternative for scenarios requiring strict data integrity.
\begin{table}[t]
\centering
\small
\setlength{\tabcolsep}{3pt}
\begin{tabular}{l|c|c|cc}
\toprule
Method & PSNR & SSIM &Angle (Right) $30^\circ$ & Distance 30cm \\
\midrule
StegaStamp & 33.07 & 0.904 & 1.27 & 1.13 \\
PIMoG & 36.06 & 0.967 & 8.27 & 6.94 \\
SRWSS & 33.98 & 0.963 & 22.97 & 12.4 \\
S2R & 42.27 & 0.962 & 6.68 & 3.51 \\
\midrule
\textbf{NiMark} & $\mathbf{\infty}$ & \textbf{1} & \textbf{0.68} & \textbf{0.46} \\
\bottomrule
\end{tabular}
\caption{Comparison with SOTA screen-shooting resilient schemes. Robustness (BER \%) vs. Visual Quality.}
\label{tab:comparison_embedding}
\end{table}

\subsection{Cross-device Robustness Evaluation}
\label{sec:cross_device}

To verify the generalization capability of NiMark across heterogeneous hardware, we conducted cross-device evaluations using a comprehensive test matrix involving three distinct display terminals (Lenovo Legion Y9000P, Envision G249G, ASUS ROG Strix SCAR Edition 8) and three different capture devices (Samsung Galaxy S20 FE, iPhone 16, Vivo S17t). This setup simulates real-world verification scenarios where the physical characteristics of screens and camera sensors vary significantly. Table \ref{tab:cross_device} reports the BER performance for each hardware combination. The consistently low BERs across all pairings demonstrate NiMark’s generalization and stability across diverse hardware environments.

\begin{table}[t]
\centering
\small
\setlength{\tabcolsep}{2pt} 

\begin{tabular}{l|cc|cc|cc}
\toprule
\multirow{2}{*}{Display Device} & \multicolumn{2}{c|}{Samsung S20 FE} & \multicolumn{2}{c|}{iPhone 16} & \multicolumn{2}{c}{Vivo S17t} \\
\cmidrule(lr){2-3} \cmidrule(lr){4-5} \cmidrule(lr){6-7}
 & BER & U-BER & BER & U-BER & BER & U-BER \\
\midrule
Lenovo Y9000P   & 0.46 & 49.41 & 0.73 & 49.43 & 0.62 & 51.89 \\
Envision G249G  & 0.37 & 48.52 & 1.31 & 51.39 & 0.72 & 49.41 \\
ASUS ROG Strix  & 1.06 & 47.92 & 0.88 & 50.80 & 0.47 & 50.31 \\
\bottomrule
\end{tabular}
\caption{Cross-device robustness evaluation (BER \% / U-BER \%). The results demonstrate generalization across different display and capture hardware combinations.}
\label{tab:cross_device}
\end{table}

\subsection{Ablation Studies}
\label{sec:ablation}
We investigate the effectiveness of key components in NiMark through controlled experiments. 
Table \ref{tab:ablation_components} summarizes the impact of architectural variants. First, replacing our SG-XOR with ``Concat+Conv'' fusion yields a near-zero U-BER, revealing that the key becomes self-contained; this explains the trivial 0\% BER, as the model bypasses image features, rendering their distortion irrelevant.
Furthermore, compared with the SG-XOR estimator, the variant using standard STE exhibits a much higher BER, as its oversimplified gradient fails to accurately model the discrete XOR logic. Finally, compared to noise layers from existing robust watermarking schemes (StegaStamp, PIMoG, SSDS, S2R), our noise layer achieves the best performance for the non-intrusive task, yielding the lowest BER of 0.46\% under real-world attacks.

\begin{table}[t]
\centering
\small
\begin{tabular}{llcc}
\toprule
Module & Variant & BER (\%) & U-BER (\%) \\
\midrule
Architecture & Cat + Conv Fusion & 0.00 & 0.00 \\
\midrule
Gradient & STE & 4.14 & 46.62 \\
\midrule
\multirow{4}{*}{Noise Layer} & StegaStamp & 2.55 & 47.64\\
& PIMoG & 1.26 & 50.53 \\
& SSDS & 1.96 & 47.36 \\
& S2R & 0.62 & 52.84 \\
\midrule
& \textbf{NiMark} & 0.46 & 49.41 \\
\bottomrule
\end{tabular}
\caption{Ablation study on Architecture, Gradient Estimator, and Noise Layer.}
\label{tab:ablation_components}
\end{table}

\subsubsection{Impact of Training Strategy}
Table \ref{tab:ablation_training} compares our two-stage strategy against single-stage baselines.
Results indicate that:
\noindent 1) Stage 1 Only: Training without the noise layer yields only limited robustness.
\noindent 2) Joint Training: While end-to-end training is feasible, simultaneously optimizing for logical binding and restoration yields suboptimal robustness compared to our progressive strategy.
\noindent 3) NiMark (S1 $\to$ S2): By decoupling the two tasks, our strategy allows the network to focus on specific objectives at each stage, achieving the best overall robustness.

\begin{table}[t]
\centering
\small
\setlength{\tabcolsep}{15pt} 
\begin{tabular}{l|c|c}
\toprule
Strategy & Description & BER (\%) \\
\midrule
S1 Only & Logic Pre-training & 4.12 \\
Joint Training & Single Stage & 0.68 \\
\midrule
\textbf{NiMark} & \textbf{Two Stage} & \textbf{0.46} \\
\bottomrule
\end{tabular}
\caption{Ablation study of training strategies. We compare our two-stage approach against logic-only pre-training and end-to-end joint training.}
\label{tab:ablation_training}
\end{table}

\section{Conclusion}
\label{sec:conclusion}

In this paper, we propose NiMark, a non-intrusive framework robust against screen-shooting. We identify the Structural Shortcut, a failure mode where networks may learn trivial identity mappings. To mitigate its impact, we introduce the SG-XOR estimator which bridges the non-differentiable gap. Furthermore, to bridge the severe domain gap caused by screen-shooting, we devise a two-stage training strategy that decouples logic learning from physical restoration. Experiments demonstrate that NiMark achieves competitive and often superior robustness compared to state-of-the-art methods while maintaining zero visual distortion. Future work will focus on real-time mobile optimization and video protection.

\clearpage

\bibliographystyle{named}
\bibliography{ijcai26}

\end{document}